\newcommand{\kms}{\mbox{km s$^{-1}$}}
\newcommand{\um}{$\mu$m}

\newcommand{\Msun}{M$_{\odot}$}

\newcommand{\Lsun}{L$_{\odot}$}

\newcommand{\cmc}{${\rm cm}^{-3}$}
\newcommand{\hii}{\mbox{$\mathrm{H\,{\scriptstyle {II}}}$}}

\newcommand{\tastar}{T$^{*}_{A}$}

\newcommand{\arcdeg}{$^o$}

\newcommand{\arcsec}{$^{''}$}

\documentclass{pasa}

\title[MALT 90: The Millimetre Astronomy Legacy Team 90 GHz Survey]{\Large MALT90: The Millimetre Astronomy Legacy Team 90 GHz Survey}
\author[Jackson et al.]{Jackson, J.M.$^{A, \dagger}$, Rathborne, J.M.$^B$, Foster, J.B.$^C$, Whitaker, J.S.$^D$, Sanhueza, P.$^A$, Claysmith, C.$^A$, Mascoop, J.L.$^A$, Wienen, M.$^E$, Breen, S.L.$^B$, Herpin, F.$^{F,G}$, Duarte-Cabral, A.$^{F,G}$, Csengeri, T.$^E$, Longmore, S.N.$^H$, Contreras, Y.$^B$, Indermuehle, B.$^B$, Barnes, P.J.$^I$, Walsh, A.J.$^J$, Cunningham, M.R.$^K$, Brooks, K.J.$^B$, Britton, T.R.$^{B,L}$, Voronkov, M.A.$^B$, Urquhart, J.S.$^E$, Alves, J.$^M$, Jordan, C.H.$^{N,B}$, Hill, T.$^{O,P}$, Hoq, S.$^A$, Finn, S.C.$^{A,Q}$, Bains, I.$^R$, Bontemps, S.$^{F,G}$, Bronfman, L.$^S$, Caswell, J.L.$^B$, Deharveng, L.$^T$, Ellingsen, S.P.$^N$, Fuller, G.A.$^U$, Garay, G.$^S$, Green, J.A.$^B$, Hindson, L.$^{V,B}$, Jones, P.A.$^{K,S}$, Lenfestey, C.$^U$, Lo, N.$^S$, Lowe, V.$^{K,B}$, Mardones, D.$^S$, Menten, K.M.$^E$, Minier, V.$^O$, Morgan, L.K.$^V$, Motte, F.$^O$, Muller, E.$^W$, Peretto, N.$^X$, Purcell, C.R.$^Y$, Schilke, P.$^Z$, Schneider-Bontemps, N.$^{F,G}$, Schuller, F.$^{ZA}$, Titmarsh, A.$^N$, Wyrowski, F.$^E$, Zavagno, A.$^T$\\
\\
{\footnotesize  
\affil{$^A$\, Institute for Astrophysical Research, Boston University, Boston, MA 02215, USA}
\affil{$^B$\, CSIRO Astronomy and Space Science, PO Box 76, Epping, NSW, 1710, Australia}
\affil{$^C$\, Yale Center for Astronomy and Astrophysics, Yale University, New Haven, CT 06520, USA}
\affil{$^D$\, Physics Department, Boston University, Boston, MA 02215, USA}
\affil{$^E$\, Max-Planck-Institut f\"{u}r Radioastronomie, Auf dem H\"{u}gel 69, D-53121 Bonn, Germany}
\affil{$^F$\, Univ. Bordeaux, LAB, UMR 5804, F-33270, Floirac, France}
\affil{$^G$\, CNRS, LAB, UMR 5804, F-33270, Floirac, France}
\affil{$^H$\, Astrophysics Research Institute, Liverpool John Moores University, Twelve Quays House, Egerton Wharf, Birkenhead CH41 1LD, UK}
\affil{$^I$\, Astronomy Department, University of Florida, Gainesville, FL 32611, USA}
\affil{$^J$\, International Centre for Radio Astronomy Research, Curtin University, GPO Box U1987, Perth WA 6845, Australia}
\affil{$^K$\, School of Physics, University of New South Wales, Sydney, NSW, 2052, Australia}
\affil{$^L$\, Dept. of Physics \& Astronomy, Macquarie University, Sydney, NSW, 2109, Australia}
\affil{$^M$\, Department of Astrophysics, University of Vienna, T\"{u}rkenschanzstrasse 17, 1180, Vienna, Austria}
\affil{$^N$\, School of Mathematics and Physics, University of Tasmania, Private Bag 37, Hobart, Tasmania 7001, Australia}
\affil{$^O$\, Laboratoire AIM Paris-Saclay, CEA/IRFU-CNRS/INSU-Universit\'{e} Paris Diderot, CEA Saclay, 91191 Gif-sur-Yvette Cedex,France}
\affil{$^P$\, Joint ALMA Observatory, Alonso de C\'{o}rdova 3107, Vitacura 763-0355, Santiago, Chile}
\affil{$^Q$\, Current address: New England College of Optometry, Boston, MA, 02115 USA}
\affil{$^R$\, Centre for Astrophysics and Supercomputing, Swinburne University of Technology, Hawthorn, Victoria 3122, Australia}
\affil{$^S$\, Departamento de Astronom\'{i}a, Universidad de Chile, Casilla 36-D, Santiago, Chile}
\affil{$^T$\, Aix Marseille Universit\'{e}, CNRS, LAM (Laboratoire d'Astrophysique de Marseille) UMR 7326, 13388, Marseille, France}
\affil{$^U$\, Jodrell Bank Centre for Astrophysics, School of Physics and Astronomy, The University of Manchester, Manchester, M13 9PL,ÊUK}
\affil{$^V$\, University of Hertfordshire, College Lane, Hatfield AL10 9AB, UK}
\affil{$^W$\, NAOJ, Chile Observatory, 2-21-1 Osawa, Mitaka, Tokyo 181-8588, Japan}
\affil{$^X$\, School of Physics \& Astronomy, Cardiff University, Cardiff CF24 3AA, UK}
\affil{$^Y$\, Sydney Institute for Astronomy (SiFA), School of Physics, The University of Sydney, NSW 2006, Australia}
\affil{$^Z$\, I. Physikalisches Institut, Universit\"{a}t zu K\"{o}ln, Z\"{u}lpicher Str. 77, 50937 K\"{o}ln, Germany}
\affil{$^{ZA}$\, European Southern Observatory, Alonso de Cordova 3107, Casilla 19001, Santiago 19, Chile}
\affil{$^\dagger$\, E-mail: jackson@bu.edu}}}
\jid{Preprint}
\doi{Accepted 2013 October 3}
\jyear{\the\year}

\usepackage[authoryear]{natbib}
\bibpunct{(}{)}{;}{a}{}{,}
\begin{document}
\begin{abstract}
The Millimetre Astronomy Legacy Team 90 GHz (MALT90) survey aims to characterise the 
physical and chemical evolution of high-mass star-forming clumps. Exploiting the unique broad frequency range and on-the-fly mapping 
capabilities of the Australia Telescope National Facility Mopra 22 m single-dish telescope\footnote{The Mopra radio telescope is part of the Australia Telescope National Facility which is funded by the Commonwealth of Australia for operation as a National Facility managed by CSIRO.} , MALT90 has obtained $3' \times 3'$ maps 
toward $\sim$2000 dense molecular clumps identified in the ATLASGAL 870\,\um\, Galactic plane survey. The clumps were 
selected to host the early stages of high-mass star formation and to span the complete range in their evolutionary states (from 
prestellar, to protostellar, and on to \hii\ regions and photodissociation regions). Because MALT90 mapped 16 lines simultaneously with excellent spatial 
(38\arcsec) and spectral (0.11\,\kms) resolution, the data reveal a wealth of information about the clump's morphologies, chemistry, 
and kinematics. In this paper we outline the survey strategy, observing mode, data reduction procedure, and highlight some early 
science results. All MALT90 raw and processed data products are available to the community. With its unprecedented large sample of clumps, 
MALT90 is the largest survey of its type ever conducted and an excellent resource for identifying interesting candidates for high resolution studies with ALMA.
\end{abstract}



\maketitle

\section{INTRODUCTION}
\label{sec:intro}

High-mass stars (M $>$ 8 \,\Msun) dominate the energy input and chemical enrichment of galaxies.  Thus, the processes by which
they form are a key component in determining the global properties and evolution of galaxies.
Although controversy remains about how high-mass stars ultimately acquire their mass,
either locally via `monolithic collapse' or from afar via `competitive accretion' (see 
Zinnecker \& Yorke 2007 and McKee \& Ostriker 2007), all theories agree that a high-mass star begins
its life in a dense core, which collapses and fragments to form one or more stars. 
(Since high-mass stars form in clusters, it is important to distinguish between molecular `clumps' that give rise to clusters 
and molecular `cores' that give rise to individual stars or close binary stellar pairs.  In this paper we refer to `clumps' as 
objects with size scales $\sim$1 pc and masses $\gtrapprox$200 \Msun. Clumps are expected to give rise to clusters or 
groups of stars.  We refer to `cores' as objects with size scales $\lessapprox$0.1 pc  and masses of $\lessapprox$100 \Msun\, 
that will form individual stars or close binary pairs.)
The dense core accretes material, and when fusion in the high-mass star begins, it enters the main-sequence 
phase while still accreting.  Eventually, the luminous, hot high-mass star terminates further star formation by disrupting the core through ionisation and outflows/stellar winds.  
These later, warmer phases are observable hot molecular cores (HMCs) and Ultra-compact (UC) \hii regions.
  
Despite these long-standing theoretical expectations, the evolution of a high-mass star remains poorly understood, primarily 
because the processes are difficult to observe. Compared with low-mass stars, high-mass stars have much shorter formation 
timescales, a much sparser Galactic distribution, and an enormous disruptive effect on their natal clouds. Observations now 
clearly demonstrate that high-mass stars indeed form exclusively in dense clumps. Consequently, new observational studies that
characterise their physical and chemical evolution are sorely needed for further progress. We aim to provide an important new 
legacy database for the study of high-mass star-formation by conducting the Millimetre Astronomy Legacy Team 90 GHz (MALT90) 
survey, a large molecular line survey of high-mass star-forming regions with the Australia Telescope National Facility (ATNF)  Mopra Telescope.

In this paper we provide an overview of the MALT90 survey and describe the motivation, telescope and observing modes, 
molecular lines selected, identification of suitable high-mass star-forming clumps, the data reduction procedure and data release.
With its unprecedented large sample of clumps, MALT90 is an important new database that will support numerous scientific
investigations for years to come. Here we highlight just a few early science results.

\section{THE SURVEY}

\subsection{Motivation}

In order to better understand the formation of high-mass stars, it is important to identify and characterise a sample of high-mass star 
forming clumps in all stages of protostellar evolution.  Fortunately, recent Galactic plane surveys of the millimetre, sub-millimetre,  
far-infrared, and mid-infrared continuum emission have identified a large number of these high-mass star forming clumps, e.g., GLIMPSE (3--8 $\mu$m, Benjamin et al. 2003; Churchwell et al. 2009), 
MIPSGAL (24 $\mu$m, Carey et al. 2005), HiGAL (70--500 $\mu$m, Molinari et al. 2010), ATLASGAL (870\,\um, Schuller et al. 2009), and
BGPS (1.1\,mm, Aguirre et al. 2011). Combined, these surveys enable us to identify and characterise a large 
number of clumps in all stages of the proposed evolutionary sequence for high-mass star-formation.

While these continuum surveys provide estimates of the dust temperatures and column densities, they have several significant limitations.  First, because the emission 
is blended along the line of sight, it is sometimes difficult to separate emission from unrelated clumps.  Moreover, continuum surveys alone cannot be used to infer distances, kinematics, or the chemical state of the material.
Fortunately, these limitations of dust continuum surveys can be overcome with complementary molecular line surveys.  By observing molecules with high 
critical densities that unambiguously trace dense gas, 
we can cleanly separate high-density, star-forming clumps from more diffuse, unrelated giant molecular clouds.  
Also, by realising that different molecular clouds  have different radial velocities, we can separate clumps that happen to lie along the same line of sight. Molecular line velocities also provide kinematic distances that delineate the clumps' Galactic distributions, and combined with the 
continuum data, allow estimates of the clump masses and luminosities.  Finally, by measuring line profiles and the intensities of multiple molecular 
species, the clumps' kinematic and chemical state can be inferred. Thus, with MALT90 we aim to determine these parameters for a large sample
of clumps identified from dust continuum surveys and in doing so characterise high-mass star-forming clumps and study their physical and chemical evolution.

With MALT90 our goal is to answer several important open questions: 
What are the statistical properties (masses, temperatures, densities, luminosities,
chemical states, and rotation rates) of these clumps? How do these properties evolve?
What is their Galactic distribution and relation to spiral arms? Can we identify rare, but
important, ``transition objects'' that represent cores in very brief evolutionary phases? How do the properties of individual star-forming
clumps relate to the large-scale properties of star-forming galaxies?

\subsection{Feasibility}

To ascertain the feasibility of such a large survey, it is critical to establish the typical line strengths, source sizes, line profiles, and 
detection rates of the clumps revealed by the various finder charts, and based on these characteristics, to choose observing 
parameters that maximise the survey's scientific usefulness. With this in mind, we conducted a MALT90 pilot survey 
(Foster et al. 2011) to test the feasibility of MALT90 and to establish the optimal observing parameters. 

The pilot observations provided critical information which guided many of our choices for MALT90, in particular that (1) Mopra 
has adequate sensitivity and speed to detect and map molecular line emission from thousands of high-mass star forming clumps in a reasonable time, 
(2) the ATLASGAL 870 $\mu$m continuum survey provides the best source list as it detects both cold and warm clumps, a highly 
desirable characteristic for studying clump evolution, (3) mapping is desirable due to the complex and varying morphologies of the 
different molecular lines, and (4) high spectral resolution is desirable due to the broad, complex line profiles often detected toward these types of clumps.

\subsection{The Mopra Telescope}

A large molecular line survey requires an efficient telescope with good sensitivity and mapping speed such as  Mopra telescope. The Mopra radio telescope  is part of the Australia Telescope National Facility which is funded by the
Commonwealth of Australia for operation as a National Facility managed by the Commonwealth Scientific and Industrial Research Organisation (CSIRO).  Mopra is a 22-m diameter telescope situated close to the Siding Spring observatory site near 
Coonabarabran, New South Wales, Australia.  It is the largest single-dish telescope operating at 3 mm  in the Southern Hemisphere.  Moreover, it is 
equipped with a sensitive, tuner-less, dual-polarization, MMIC receiver and a broad-bandwidth, flexible backend (the Mopra Spectrometer, 
MOPS)\footnote{The University of New South Wales Digital Filter Bank used for the observations with the Mopra Telescope was provided with support from the Australian Research Council.} which allows for the processing of an 8-GHz wide bandwidth. MOPS can operate in two modes, a `wideband' mode that produces 
an 8-GHz spectrum anywhere within the 3-mm band, and a `zoom' mode that simultaneously produces 16 spectra, each with a bandwidth 
of 137.5 MHz and with fixed central frequencies selected  by the user.   At 90 GHz, the spectral resolution is 3.3 \kms\ in wideband mode and 
0.11\,\kms\ in zoom mode. For typical observing conditions, the system temperatures near 90 GHz are $\sim$200 K.  The antenna's 
main-beam efficiency at 90 GHz is 0.49 (Ladd et al. 2005). Because of its ability to detect multiple spectral lines over a broad range of frequencies, Mopra has been used extensively to conduct spectral surveys of star forming regions.  Of particular note are the pioneering studies of  Purcell et al. (2006, 2009) which conducted  spectral surveys before substantial improvements allowed more efficient observations.
 
Using the On-the-Fly (OTF) mode, the Mopra telescope can also efficiently image the spatial distribution of spectral lines by scanning the telescope back and forth across the sky 
in a raster pattern, and collecting data as it scans.  Because the `off' position used for sky-subtraction is shared for all positions in a raster row, 
the time for slewing is minimised and the time for on-source integration is maximised in the OTF mode.  Redundancy is achieved by oversampling the beam 
in the scan direction (collecting data every 2 s) and overlapping subsequent rows.  Thus, the maps are not only efficient; they are also of high 
quality. As such, Mopra is an excellent survey instrument that can simultaneously image 16 spectral lines at high spectral resolution.  
For MALT90 we use Mopra in the OTF mode with MOPS in the zoom mode configured to a central frequency of $\sim$90 GHz.

\subsection{Line Selection}

We choose the 90 GHz portion of the spectrum because it is rich in diagnostic lines.
Many molecules with large dipole moments have rotational transitions to their ground or their first rotationally excited states near 90 GHz.
Since these transitions have  large Einstein A coefficients, they require high densities for excitation (typically n$>$10$^{5}$\,\cmc).  Such high densities are found in dense star-forming clumps, but not in the far more diffuse giant molecule clouds typically traced by CO.  Thus,  the 90 GHz lines from high dipole moment molecules are unambiguous probes of the dense regions that will give rise to the stars.  Diffuse gas from unrelated giant molecular clouds in their vicinity or along the line of sight simply have insufficient density to excite these lines.

Also, since the high dipole moment molecular lines near 90 GHz span a large range of excitation
energies (from $\sim$5 to 200 K) and critical densities (from $\sim$10$^{5}$ to 10$^{6}$\,\cmc), they indicate distinct physical
conditions within cores. Furthermore, because the chemistry evolves dramatically from
simple species in cold clumps to quite complicated species in hot regions, the 90 GHz lines also probe
distinct stages of chemical evolution. Finally, because the molecular line observations supply
velocities, they provide kinematic information (e.g., rotation, infall, turbulence, and
outflows) that continuum surveys cannot.  Indeed, by comparing the observed LSR velocities with a model of Galactic rotation, kinematic distances to the star forming clumps can be estimated.

The most important diagnostic lines in the 86 to 93 GHz spectrum are the density tracing, bright, ground state transitions of the abundant molecular 
species: N$_2$H$^+$ (1--0), HCO$^+$ (1--0), HCN (1--0), and HNC (1--0).  After CO and its isotopologues, these are among the brightest molecular lines in the 3 mm 
band. These lines can be used to reveal the physical and chemical state of the gas.  For instance, N$_2$H$^+$ is thought to be chemically robust in colder gas because its precursor molecule N$_2$ 
resists freezing onto dust grains, and HNC is enhanced in very cold gas (e.g., Bergin \& Tafalla 2007).  In addition to these high density tracers, we have 
also selected some rarer molecular lines with isotopic substitutions which are usually optically thin, such as $^{13}$CS, $^{13}$C$^{34}$S, H$^{13}$CO$^+$ and H$^{13}$CN.  
These lines can be used to estimate optical depths and hence column densities.  They are particularly useful to understand self-absorbed line profiles in the 
optically thick main lines, for they reveal the true systemic velocity and the velocity of the peak column density for each clump.  

Another group of selected molecules requires high densities and temperatures for their formation and/or excitation.  These molecules, CH$_3$CN, HC$_3$N, HC$^{13}$CCN, and HNCO, are typically detected only toward warm, 
dense hot clumps associated with the protostars and \hii\ regions.  Thus, they can be used as an independent check on 
the evolutionary state.  We also include three diagnostic lines that arise only under specific conditions.  The first is the H41$\alpha$ recombination line, which 
arises only in dense \hii\ regions.  The second, SiO (2--1), is formed when SiO is formed from silicate-bearing grains that are destroyed by shocks.  Consequently, SiO lines are unambiguous 
tracers of shocks; usually these arise from protostellar or stellar outflows.  Finally, the C$_2$H lines near 87 GHz are thought to be distinct chemical tracers of the ionized/molecular gas interface regions 
called photodissociation regions or PDRs.  Strong C$_2$H emission thus signals that a high-mass protostar has left the protostellar and entered the \hii\ region phase. 

We note that molecular chemistry is often complex, and the molecular tracers described above may not be unambiguous tracers in all situations.  For example, HNCO, in addition to probing hot cores, may also be enhanced in shocks (Rodriguez-Fern\'andez et al. (2010).  Moreover, C$_2$H has been found to trace both early (protostellar) and late (PDR) stages of high-mass star formation (Beuther et al. 2008).  Indeed, Sanhueza et al. (2013) find compact C$_2$H emission from a very cold infrared dark cloud with no obvious PDR. Moreover, HC$_3$N and C$_2$H emit strongly in the cold, low-mass, star-forming cores in Taurus (e.g., Pratap et al. 1997).  With its large sample of clumps and the variety of its observed lines, MALT90 can help determine exactly what the various molecular species are tracing chemically.

\subsection{Identifying star-forming clumps}

Due to the small solid angle occupied by star-forming clumps emitting strongly at 90 GHz, and the relative faintness of these lines compared with CO lines, a blind, fully-sampled survey of several square degrees 
or more of the Galactic plane is impractical.  Thus, to survey significant numbers of star-forming clumps efficiently, we use the
APEX Telescope Large Survey of the Galaxy (ATLASGAL) 870\,\um\, survey (Schuller et al. 2009), the first complete sub-mm inner Galactic
plane survey, to select dense, star-forming clumps. ATLASGAL is ideal for this purpose because it detects both cold
prestellar clumps as well as warm clumps that already contain protostars, \hii\, regions, and hot cores.  From the ATLASGAL sources, we restrict our range of Galactic longitudes to $-60$ deg to $+20$ deg.

To estimate the completeness of the MALT90 survey to high-mass star-forming clumps, we must first estimate the maximum distance at 
which ATLASGAL can detect such a clump.  Since high-mass stars form in clusters, the highest mass clumps detected by 
ATLASGAL will form not single stars, but rather star clusters. We estimate the mass of the most massive star formed in a cluster as follows.  
First, we assume that clumps will convert their mass into stars with an efficiency of $\sim$30\% (e.g., Alves et al. 2007). Although this efficiency is estimated for low-mass star-forming regions, we assume that the same efficiency also applies to high-mass regions.   Next, we apply the 
empirical relationship between cluster mass and the mass $M^*$ of the most massive star, $M^* \sim 1.2 M_{cluster}^{0.45}$ (Larson 2003).  
Using this procedure, we find that clumps with masses $M_{clump} > 200$ \Msun\ are likely to form a high-mass star with $M^* > 8$ \Msun. 

Due to the high sensitivity of the ATLASGAL survey, such $M_{clump} > 200$ \Msun\ high-mass star forming clumps can be detected at fairly 
large distances in the Galaxy.  Making the usual assumptions about dust temperature, dust emissivity, and gas-to-dust mass ratios, one can 
estimate the mass of dense clumps from sub-mm continuum data to within factors of a few.  Schuller et al. (2009) estimate that, for cold dust with $T_{dust} = 10$ K, the 
ATLASGAL 5$\sigma$ 870 $\mu$m flux sensitivity of 0.25 Jy corresponds to a clump mass of 200 \Msun\ at a distance of 10 kpc.  Thus, 
ATLASGAL detects all cold, high-mass star-forming clumps out to a distance of 10 kpc.

This 0.25 Jy flux limit is appropriate for cold dust clumps.  Once a high-mass protostellar clump leaves its prestellar phase, however, the
dust will become heated and more luminous in the sub-mm.  Thus, for the same mass, a clump containing a protostar or \hii\ region would 
have a correspondingly higher flux.  For instance, as the dust temperature increases from 10 K to 20 K, the 870 $\mu$m flux increases 
by a factor of $\sim$2 for the same mass clump.  In other words, the warmer high-mass star-forming clumps will be more luminous, or, 
equivalently, will be seen at even larger distances. We used the flux limit  of 0.25 Jy to select the clumps to observe for MALT90.
 
\subsection{Clump classification}

\begin{figure}
\begin{center}
\includegraphics[scale=0.29, angle=0]{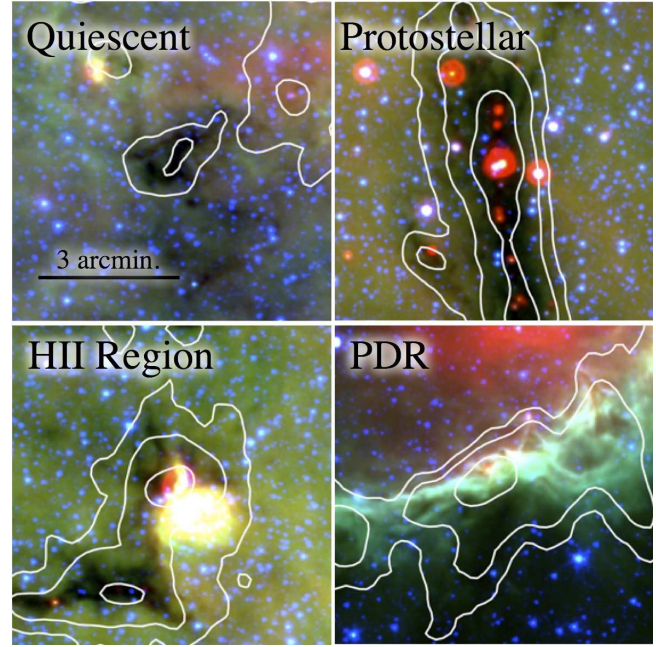}
\caption{The evolutionary classification scheme of clumps based on {\it Spitzer} images.  See text for a complete description. The images show {\it Spitzer} images from the GLIMPSE and MIPSGAL surveys.  In each image the blue, green, and red colours correspond to 3.6 $\mu$m, 8.0 $\mu$m, and 24 $\mu$m, respectively. The white contours indicate 870 $\mu$m emission from the ATLASGAL survey (Schuller et al. 2009).}
\label{classification}
\vspace{-0.8cm}
\end{center}
\end{figure}

Although ATLASGAL is an excellent finding chart for dense clumps, it contains a heterogeneous mix of warm and cold clumps spanning a 
wide range of masses. In order to adequately sample high-mass star forming dense clumps in both the colder prestellar phase and the 
warmer protostellar and \hii\ region phases, it is necessary to classify the evolutionary state of the star-formation activity taking 
place within the clumps.  Fortunately, the {\it {Spitzer}} infrared surveys GLIMPSE and MIPSGAL allow us to broadly separate the clumps into 
the various evolutionary states based on their IR characteristics. We use these data to categorize the clumps into four distinct evolutionary groups (see Fig. ~\ref{classification}):

\begin{itemize}
\item Prestellar:  Since they are cold and very dense, prestellar cores will be infrared dark in the GLIMPSE 3.6 to 
8.0\,\um\,  and MIPSGAL 24\,\um\, images.
\item Protostellar:  Accretion onto a protostar is signalled by warm dust.  Thus, an embedded 
24\,\um\, point source in MIPSGAL indicates an embedded, accreting protostar 
(\citealp{Chambers09}). These are often also associated with extended, green emission in 
GLIMPSE 3.6 to 8.0\,\um\, three-colour images from an excess of 4.5\,\um\, emission (`green fuzzies'). This emission likely traces shocked gas 
due to outflows in the immediate vicinity of a newly-forming star.
\item \hii\, region:  When the \hii\, region forms, the surrounding gas and dust are heated and the infrared emission 
associated with the core will be bright in both GLIMPSE and MIPSGAL images. Moreover the PAH feature that dominates the {\it Spitzer} 8 
$\mu$m band is typically bright due to fluorescent excitation.  The morphologies of \hii\, regions 
range from compact to extended, with some displaying bubble-like morphologies.  
\item PDR: The UV radiation from a recently formed high-mass star  will produce a photodissociation region (PDR) 
at the molecular/ionised gas interface.  The PDR will be extended and emit strongly in the GLIMPSE 8\,\um\, images.  
Moreover, extended 24\,\um\, emission indicative of hot dust will also be present. In practice, we classify soures as PDRs when they have very extended 8 $\mu$m morphologies and no readily identifiable compact \hii\ region.
\end{itemize}

With several thousand compact sources identified within the ATLASGAL survey
 (e.g., Contreras et al. 2013) and the limited observing time allotted to the MALT90 survey, we chose to observe randomly select sources from the ATLASGAL compact source catalog.
An ATLASGAL source was first classified using the above method before being placed within the queue of 
sources to be observed. For a particular time within any given observing shift, sources were selected from this queue based on 
their classification, peak flux, and elevation. Our aim was to observe a large sample of clumps at all flux levels and in each of the various evolutionary 
stages.

\subsection{Observing Procedure}

Because the MALT90 survey aims to map and characterise a large number of
sources and compare their physical and chemical properties, all clumps were observed
using the same mapping parameters and in the same molecular transitions.

\begin{table*}
\small
\begin{center}
\caption{Spectral Lines in the MALT90 Survey}\label{malt90-lines}
\begin{tabular}{lcl}
\hline Transition & Frequency (MHz) & Tracer \\
\hline 
 N$_{2}$H$^{+}$   (1--0)	& 93173.772  & Density, chemically robust \\
 $^{13}$CS (2--1)	& 92494.303  & Optical depth, Column density, V$_{LSR}$\\
 H41 $\alpha$            & 92034.475 & Ionized gas\\
 CH$_{3}$CN 5(0)--4(0) 	& 91987.086  & Hot core\\
 HC$_{3}$N (10--9 )	& 90978.989  & Hot core\\
 $^{13}$C$^{34}$S (2--1)  & 90926.036 & Optical depth, Column density, V$_{LSR}$\\
 HNC (1--0) 	        & 90663.572  & Density; Cold chemistry\\
 HC$^{13}$CCN (10--9)	& 90593.059  & Hot core\\
 HCO$^{+}$ (1--0)	& 89188.526  & Density; Kinematics\\
 HCN (1--0)	        & 88631.847  & Density\\
 HNCO  4(1,3)--3(1,2)  	& 88239.027  & Hot core\\
 HNCO  4(0,4)--3(0,3)	& 87925.238  & Hot core\\
 C$_{2}$H (1--0) 3/2--1/2 & 87316.925 & Photodissociation region\\
 HN$^{13}$C (1--0)	& 87090.859  & Optical depth, Column density, V$_{LSR}$\\
 SiO 	(1--0)          & 86847.010  & Shock/outflow\\
 H$^{13}$CO$^{+}$ (1--0) & 86754.330  & Optical depth, Column density, V$_{LSR}$\\
\hline
\end{tabular}
\medskip\\
\end{center}
\end{table*}

The observations were conducted over three Austral winter observing seasons: July 11 to September 21 2010, May 7 to September 29 2011, and June 29 to October 31 2012.
Over these three seasons more than 2,000 hours of observing time was assigned to the MALT90 project in blocks of time we shall call `sessions.'  Each session was about 14 hours 
long for the first two years and 11 hours for the third year. To begin each session the telescope pointing was checked using a  SiO maser with an accurately determined position (Indermuhle et al., 2013), followed by a short position-switched 
observation of a bright \hii\ region (almost always G300.968+01.145). Because the G300.968+0.1.145 \hii\ region contains strong emission from most of the selected molecular 
transitions, these data were used to check quickly that the telescope, receiver, and backend had been configured correctly.  Moreover, since the source was observed at the 
beginning of almost every observing session, the repeatability of these measurements provides an important assessment of the system reliability, performance, and stability. 
Based on these repeated observations of G300.968+01.145 , Foster et al. (2013) conclude that the accuracy of telescope pointing is a function of elevation and that the system gain varies over the course of each season, 
with both source elevation and the time of day contributing to the gain variations.  Nevertheless, these variations are small, and the pointing is accurate to 8\arcsec\ and system gain is repeatable to within about 30\%. At frequencies near 90 GHz the Mopra beam size is 38\arcsec\ FWHM.

Once the observer confirmed that the brightest lines in G300.968+01.145  were detected and that their intensities and profiles were similar to a previous `template' 
observation, the observer proceeded to observe the session's first target clump. Before mapping the first clump, the telescope pointing was checked on a nearby 
SiO maser (Indermuhle et al., 2013), typically within 30\arcdeg.  Once the pointing was confirmed, or if necessary, adjusted, the source mapping commenced. This procedure (a pointing check followed by mapping) was repeated for subsequent sources. At any given time during the shift, a source was selected to be observed 
based on its elevation. A restriction of 35\arcdeg\ $<$ El $<$ 70\arcdeg\ was implemented to avoid regions where the telescope pointing model is less reliable and the scanning less efficient.

We obtained two $3' \times 3'$ maps around each ATLASGAL clump.  Each map was obtained using OTF mode, scanning in Galactic coordinates. To facilitate the removal of any 
striping or baseline variations in the maps that result from fluctuations in the sky opacity, two maps were obtained for each clump by scanning in orthogonal directions: one in 
Galactic Latitude and the other in Galactic Longitude. Each map took $\sim$31 min to complete. At the end of each scanning row a reference position was observed. For 
each map, the reference position was offset from the clump position $+1$ degree in Galactic Latitude for sources with $b > 0$ and $-1$ degree for sources with $b < 0$. 
The system temperature was determined by measuring the emission from an ambient and a cold `paddle' every 15 minutes (twice per map). 

For all observations, the `zoom mode' of the MOPS spectrometer was configured to a central frequency of 89.690 GHz.  The zoom mode allows the simultaneous observation of 16 individual IFs within a range of 8 GHz. Each IF was 137.5 MHz wide, with 4096 spectral channels, corresponding to a velocity resolution of 0.11 \kms. We  selected
16  IFs to include the specific lines listed in Table~\ref{malt90-lines}. 
     
Over all the observing seasons MALT90 completed 2,014 maps which covered over 2,014 ATLASGAL sources (some maps contain more than one ATLASGAL source).  Roughly equal numbers of prestellar clumps, protostellar clumps, \hii\, 
regions, and PDRs  were observed.  

\subsection{Data Reduction}

The data reduction for each map was performed using an automated python-based pipeline (using the ATNF packages ASAP, Livedata, and Gridzilla). To minimise artefacts due 
to noise spikes and baseline ripples in the spectrum of the reference position, we modified Livedata to smooth the reference spectrum with an 11-channel Hanning 
smooth before each reference subtraction.  This procedure is valid because baseline variations are slowly-varying in frequency.  After removal of the bandpass shape with 
the smoothed reference-position spectrum, the individual on-source spectra were co-added and gridded onto a uniform 9$''$ grid.  To improve pixel-to-pixel variation and to eliminate gaps in the maps, we applied a 12$''$ 
Gaussian spatial smooth to the data.  This degrades the angular resolution from about 36$''$ to 38$''$.   The orthogonal maps were then combined using a system temperature 
weighting to produce a single data cube per source for each of the 16 IFs. After gridding, the integration time per pixel in the maps is equivalent to $\sim$30 secs. 

\begin{figure}[h]
\vspace{-0.55cm}
\begin{center}
\includegraphics[scale=0.44, angle=0]{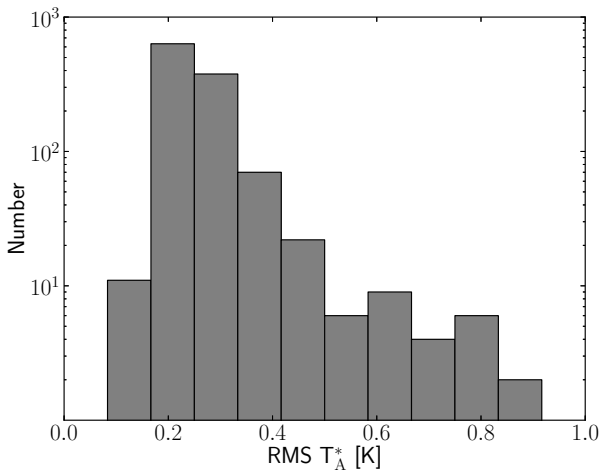}
\caption{ Histogram of the measured rms noise in \tastar\ per channel for
 the first three years of MALT90 (2,014 sources). The typical rms noise
 is $\sim$\tastar\ = 0.25 K per 0.11 \kms\ channel.}
\label{malt90noise}
\end{center}
\vspace{-0.82cm}
\end{figure}

Given that the typical system temperatures during the observations were 180 to 300 K, the typical rms noise is $\sim$\tastar\ = 0.25 K per 0.11 \kms\ channel. A histogram 
of the measured rms noise in \tastar\  is presented in Figure~\ref{malt90noise}. All spectra are presented on the antenna temperature (\tastar) scale, corrected to the top of the 
atmosphere. To convert to main-beam brightness temperatures, one should divide the antenna temperatures by the main-beam efficiency of 0.49 \citep{Ladd05}.  For extended sources $>80''$, sources begin to couple to the inner error beam.  In this case a more appropriate efficiency would be 0.65 (Ladd et. al 2005).  

\subsection{Data release}

The MALT90 dataset is an important new database that will support numerous scientific
investigations for years to come. To support these community efforts, the MALT90 data are freely available in annual data
releases. MALT90 data were released with the support of the ATNF (http://atoa.atnf.csiro.au/MALT90) shortly after each of the
observing season finished and data quality checks were complete. We have built a fully automated
data reduction pipeline, taking data straight from the telescope and producing publication quality
images and processed cubes. Currently released data products include the raw data, processed data cubes, and
moment maps. We will also make available (via the MALT90 web page, http://malt90.bu.edu/) a large database 
containing additional custom products such as N$_{2}$H$^{+}$ opacities, line parameters determined from Gaussian fits to 
detected lines, and derived kinematic distances.

\begin{table}
\small
\begin{center}
\caption{Parameters of the MALT90 Survey}\label{obs-param}
\begin{tabular}{lc}
\hline Parameter & Value \\
\hline 
Galactic Longitude Range	                      &  $-60$ to 20 deg\\
Galactic Latitude Range		& $-1$ to $+1$ deg\\
Total Number of Maps (to date)  & 2,014\\
 Frequency Range                      & 86.7 - 93.2 GHz \\
 Number of Spectral Lines (IFs)       & 16 \\	   
 Angular Resolution   (FWHM)                & 38$''$ \\
 Spectral Resolution                  & 0.11 \kms \\
 Typical Noise (\tastar) per channel  &  0.25 K \\
 Typical T$_{\rm sys}$                  & $180 - 300$ K  \\
 Effective Integration Time per Pixel &  30 s \\
 Telescope Main Beam Efficiency       & 0.49 \\
 Map Size (pixels)                             &  $27 \times 27$ \\ 
 Map Size (angle)			& $3' \times 3'$\\
Pixel Size                           &  $9''$ \\
 
\hline
\end{tabular}
\medskip\\
\end{center}
\end{table}

\section{SCIENCE WITH MALT90}

The MALT90 dataset represents a vast database for which numerous scientific
investigations can be conducted. Our choice of ATLASGAL as the basic finding chart has
proven to be a good one; MALT90 detected every ATLASGAL core in at least one 90
GHz molecular line. The brightest four lines, HCN, HNC, N$_2$H$^+$, and HCO$^+$, are typically
detected toward all clumps, and other key diagnostic lines, e.g., hot core tracers or isotopic
lines, are usually detected toward the warmest, most luminous clumps. There are
numerous MALT90 scientific investigations underway; here we present preliminary results from only a few.

\subsection{Determining distances and Galactic structure} 

\begin{figure*}
\vspace{-2.0cm}
\begin{center}
\includegraphics[scale=0.7, angle=0]{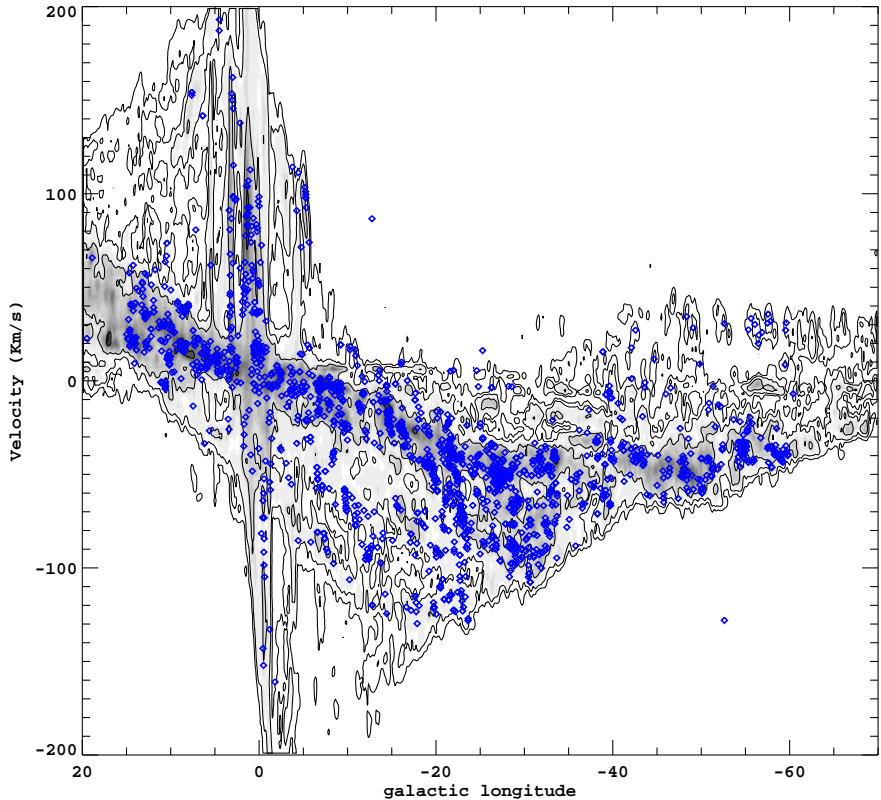}
\vspace{-1.3cm}
\caption{A longitude-velocity diagram of the MALT90 region.  The gray scale is the CO (1-0) data from Dame et al. (2001).  Crosses represent the position and velocity derived from averaging the central 9 pixels of each MALT90 map.  Velocities are determined from integrated-intensity weighted fits of HCN, HCO$^+$, N$_2$H$^+$, and HNC (1-0) (Whitaker et al., in prep.).}
\label{lvco}
\end{center}
\vspace{-0.55cm}
\end{figure*}

Distance is a key parameter to understanding the properties of high-mass star-forming dense clumps.  Without distances, 
it is impossible to estimate such important physical quantities as mass, luminosity, and Galactic distribution.  Consequently, one of the 
prime science goals for MALT90 is to determine distances to each of the target clumps. Fortunately, by measuring the molecular 
line velocities in the MALT90 data, kinematic distances to the molecular clumps can be estimated. Because distances to the emission
detected via dust continuum surveys such as the ATLASGAL, BGPS, and HiGAL are unknown, MALT90 can overcome this limitation 
by providing distances to thousands of clumps.  The accuracy of kinematic distances depends on the accuracy of the rotation curve, the degree to which non-circular motions contribute to the measured LSR velocities, and the resolution of the near/far kinematic distance ambiguity (Whitaker et al., in prep.).    Nevertheless,the kinematic distance method remains the only practical means to estimate distances to thousands of star-forming regions.

Once the distances are determined, the Galactic distribution can be obtained.  Since high-mass stars form in spiral arms, 
but typically move $< 100$ pc from their birthplaces during their short lifetimes, high-mass star forming clumps should delineate 
the Milky Way's spiral structure. Thus, MALT90 should prove to be a valuable dataset for determining Galactic structure. A detailed 
analysis of the distances to the MALT90 targets and their Galactic distribution will be presented by Whitaker et al. (in preparation).  
Here we present a preliminary analysis of the data from the first and 
second years. 

Radial velocities have been extracted for MALT90 clumps observed in the first and second seasons (Figure~\ref{lvco}). 
To determine a systemic velocity for a MALT90 source, we calculated the integrated-intensity-weighted mean of the velocities from Gaussian fits to the most prominent peaks in the HCO+, HNC, N2H+, and HCN spectra averaged over the center nine pixels of the map.   To reduce sensitivity to fitting difficulties, if at least two other lines had significant detections then the HCN velocity was excluded from the calculation due to complications of self-absorption and overlapping hyperfine lines.  Details can be found in Whitaker et al. (in prep.).
 
\begin{figure*}
\vspace{-0.6cm}
\begin{center}
\includegraphics[scale=1.0, angle=0]{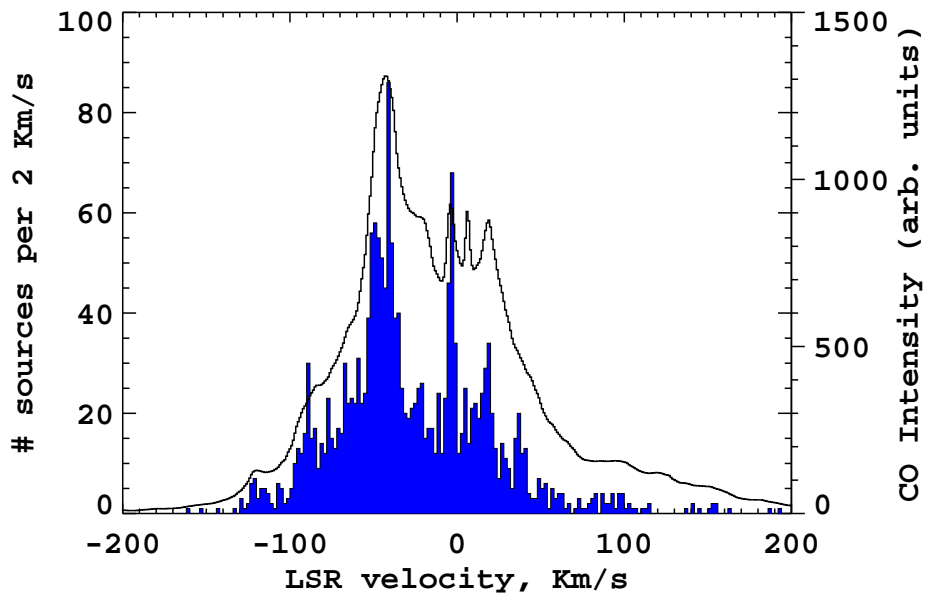}
\vspace{-0.1cm}
\caption{A histogram of the number of MALT90 sources that have a particular LSR velocity.  The black spectrum in the background represents the CO (1-0) spectrum obtained by averaging together all positions in the Columbia-CFA CO survey (Dame et al. 2001) in the MALT 90 region.  The sharp concentration of sources at particular LSR velocities indicate Galactic spiral arm structure (Whitaker et al., in prep.). }
\label{velco}
\end{center}
\vspace{-0.6cm}
\end{figure*}

A comparison of these 
velocities with CO data from the CfA-Columbia survey (Dame et al. 2001) shows that the dense clumps are largely confined to the 
brightest CO emission features, such as the `molecular ring.' Nevertheless, the distribution of dense clump velocities differs significantly 
from the CO spectrum averaged over the surveyed region (Figure~\ref{velco}).  The distribution of the  dense clumps' velocities has sharp peaks near $-$5, $-$40, and $-$52 \kms, whereas the CO spectrum is much smoother.  The simplest interpretation is that the 
dense clumps are confined to spiral arms, but the CO is more widespread throughout the Galactic disk.  

\subsection{Chemical variation}

\begin{figure*}
\vspace{-0.1cm}
\begin{center}
\includegraphics[scale=0.6, angle=0]{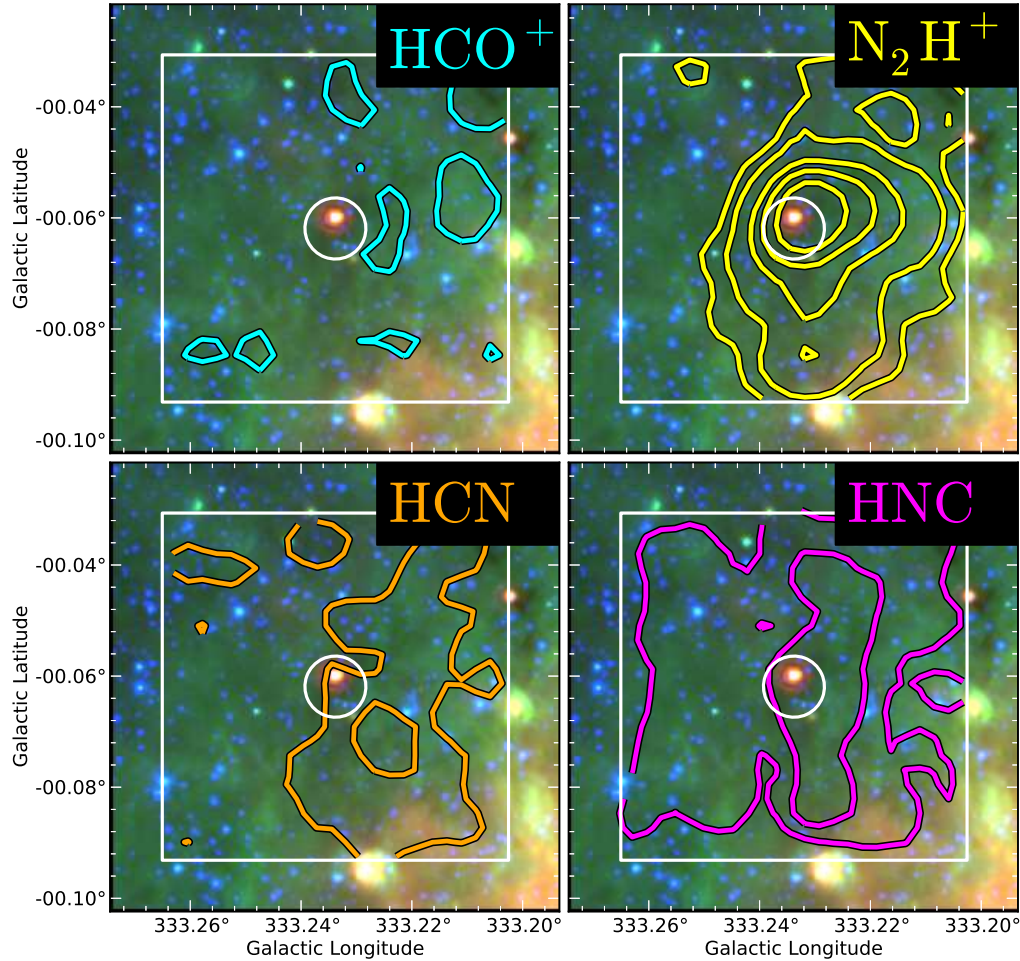}
\caption{MALT90 images of the ATLASGAL continuum source G333.234-00.062.  The color image are GLIMPSE/MIPSGAL 3-colour images with blue = 3.6 $\mu$m, green = 8.0 $\mu$m, and red = 24 $\mu$m.  The contours represent integrated intensity molecular line emission from MALT90, with HCO$^+$ (1-0) in cyan (upper left), N$_2$H$^+$ (1-0) in yellow (upper right), HCN (1-0) in orange (lower left), and HNC (1-0) in pink (lower right).  Contour levels are drawn at signal-to-noise levels of
1.5, 3, 7, 11, and 17.  This is an example of a source with strong N$_2$H$^+$ emission, but weak or absent HCO$^+$, HCN, and HNC emission.  Since the maps were taken simultaneously, the variations in intensity are real and not due to calibration errors. The circle represents the Mopra FWHM beam.
}
\label{n2hprich}
\end{center}
\vspace{-0.5cm}
\end{figure*}

\begin{figure*}
\vspace{-0.1cm}
\begin{center}
\includegraphics[scale=0.6, angle=0]{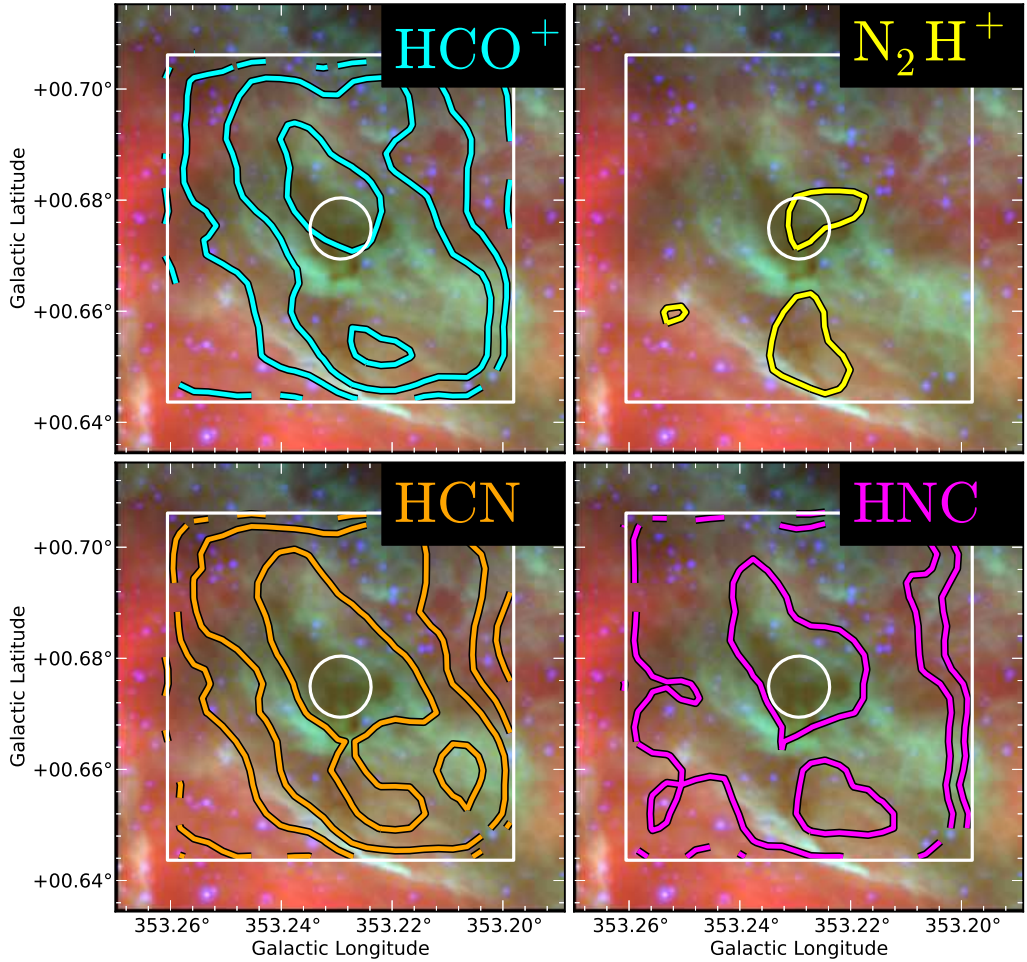}
\caption{MALT90 images of the ATLASGAL continuum source G353.229+00.675.  The color image are GLIMPSE/MIPSGAL 3-colour images with blue = 3.6 $\mu$m, green = 8.0 $\mu$m, and red = 24 $\mu$m.  The contours represent integrated intensity molecular line emission from MALT90, with HCO$^+$ (1-0) in cyan (upper left), N$_2$H$^+$ (1-0)  in yellow (upper right), HCN (1-0) in orange (lower left), and HNC (1-0) in pink (lower right). Contour levels are drawn at signal-to-noise levels of
1.5, 3, 7, 11, and 17. This is an example of a source with weak or absent N$_2$H$^+$ emission, but strong HCO$^+$, HCN, and HNC emission.  Since the maps were taken simultaneously, the variations in intensity are real and not due to calibration errors.  The circle represents the Mopra FWHM beam. }
\label{n2hppoor}
\end{center}
\vspace{-0.5cm}
\end{figure*}

A second key goal of the MALT90 survey is to investigate the chemical evolution of dense clumps.  Chemical models of clumps suggest 
that the abundances of different molecular species change dramatically with time, and thus, can potentially indicate the clump's evolutionary 
phase (e.g., Sanhueza et al. 2012).  An analysis of the first year's MALT90 data finds significant chemical evolution of the star-forming clumps (Hoq et al. 2013).  
A particularly surprising result shows interesting variations in the N$_2$H$^+$ abundances.  For some sources, the HCO$^+$, HCN, 
and HNC(1-0)  lines are strong, but the N$_2$H$^+$ (1-0) line is significantly weaker.  For others, however, the N$_2$H$^+$ (1-0) line is strong, but 
the HCO$^+$, HCN, and HNC (1-0) lines are weaker.  Two exmples of these chemically anomalous sources are shown in 
Figures~\ref{n2hprich}~and~\ref{n2hppoor}.   Although optical depth effects may account for these apparently chemically extreme sources, in at least some cases it is clear that the chemical abundances among these four molecular species vary significantly.  The reader is referred to Hoq et al. (2013) for details.  Barnes et al. (2013) have also found significant HCO$^+$/N$_2$H$^+$ variation in high-mass star-forming regions which they attribute to  evolutionary effects as the clumps form massive stars.

\subsection{Comparison with Extragalactic Observations}  

Understanding Galactic high-mass star formation is critical for interpreting observations of galaxies.  Luminous high-mass stars dominate 
the stellar population observable at large distances.  Moreover, the early universe was characterized by intense high-mass 
star formation (e.g., Hughes et al. 1998). Star formation in external galaxies is often characterized by the `Kennicutt-Schmidt law' 
(Kennicutt 1998; Schmidt 1959), an empirical correlation between the surface gas-mass density of a galaxy and the local star 
formation rate, typically averaged over at least 100 pc scales.  On several kpc scales, comparable to the size of galactic disks, 
a correlation is found between the far-IR luminosity (a measure of star formation rate) and the luminosity of CO line emission 
(a measure of molecular gas content). Gao \& Solomon (2004) found an even better correlation between IR luminosity and HCN 
luminosity.  Because HCN has a higher dipole moment than CO, it requires denser gas ($n(H_2) > 3 \times 10^4$ cm$^{-3}$) to excite
its $J = 1 - 0$ transition. 
Gao \& Solomon (2004) propose that the better correlation with HCN than with CO arises because star formation occurs 
only in the dense gas, which HCN traces better than CO.  More diffuse gas traced by CO does not directly participate in star formation, 
and thus is more poorly correlated with FIR emission.  We aim to understand the physical origin for this HCN-IR correlation by studying 
it in individual dense clumps with MALT90.  We can then hope to understand the basis for the HCN-IR correlation, and under what 
conditions and size scales it should be expected to hold.  Since the Schmidt-Kennicutt law is the usual prescription for star formation 
in cosmological simulations, it is important to understand its origin and applicability.
     
Recent studies have extended this extragalactic relationship down to the scales of molecular clumps. Wu et al. (2010) mapped 50 dense clumps (mostly \hii\, regions) in HCN and verified that the Gao \& Solomon extragalactic HCN-IR relation also holds for the clumps. Their data hint at a deviation from the linear relation at the lowest luminosities, but this portion of the relation is poorly sampled. The Census of High- and Medium-mass Protostars (CHaMP; Barnes et al. 2011) data, Ma et al. (2013) provide measurements for an additional 303 dense clumps. This larger sample also falls on roughly the same relationship as the extragalactic sample, although again there is a deviation at low luminosity. 

\begin{figure}
\vspace{-0.6cm}
\begin{center}
\includegraphics[scale=0.34, angle=0]{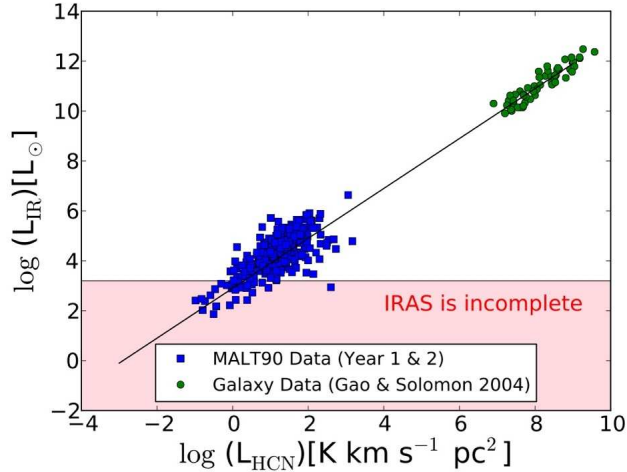}
\caption{A plot of HCN luminosity versus far-infrared continuum luminosity.  The points at the upper right (green) are for a sample of galaxies reported in Gao \& Solomon (2004).  The points at the lower left (blue) are for individual clumps observed with MALT90.  We include only those clumps which correspond to a catalogued IRAS source.  The analysis for both the clumps and the galaxies is identical. The line shows the fit from Gao \& Solomon (2004).}
\label{HCN-IR}
\end{center}
\vspace{-0.7cm}
\end{figure}

MALT90 provides a valuable addition to the number of Galactic dense clumps mapped in HCN. In the first two seasons' data there are 265 MALT90 clumps with solid detections in IRAS. Using this set of clumps we derived HCN luminosities and IR luminosities using the prescription of Gao \& Solomon (2004). Figure~\ref{HCN-IR} shows the HCN vs. FIR luminosities for the MALT90 clumps.  The MALT90 results confirm that, for these sources, the extragalactic HCN-IR relation holds over 10 orders of magnitude, and extends to individual dense clumps.  The simplest interpretation is that the basic units of star formation are dense clumps, which have a characteristic HCN/IR ratio.  Hence, galaxies, which contain a large number of clumps, will have the same HCN/IR ratio, and the global HCN and IR luminosities simply reflect the total number of clumps contained by the galaxy.

This simple interpretation must be verified, however, and numerous caveats bear investigation.  For example, IRAS lacks the sensitivity and angular resolution required to detect IR luminosities below 150 \Lsun\ for the typical MALT90 clump distance of 5 kpc.  Indeed, toward many of our less luminous HCN clumps there are no IRAS sources detected.  These sources are not plotted in Figure~\ref{HCN-IR}.  Fortunately, the increased sensitivity of the Herschel HiGAL survey will allow us to derive more accurate bolometric luminosities for the fainter sources. In fact, at low luminosities we expect a deviation from the linear HCN-IR relationship, because many of our cold prestellar clumps are detected in HCN but have no protostars detected by {\it {Spitzer}}. There is also some evidence that the HCN-IR relationship is not constant in extragalactic systems. For instance, the giant molecular clouds in M33 have significantly less HCN than would be expected from the Gao \& Solomon relation (Rosolowsky et al. 2011).

In the future, we will investigate the HCN-IR correlation as a function of evolutionary state and of bolometric luminosity.  We will also investigate whether other molecules such as HNC, N$_2$H$^+$, or HCO$^+$ show a tighter correlation, and whether the correlation improves when HiGAL bolometric luminosities are used.   If so, these molecules may be more useful probes of the dense gas and star formation content of external galaxies.  Our ultimate goal is to understand why the extragalactic HCN-IR relationship exists, and how to interpret these global  measurements in the context of clump properties.  

\subsection{The Link to ALMA}

MALT90 will be valuable not only in its own right, but also as the definitive finding chart of high-mass clumps to identify key ALMA targets.   
With its unprecedented large sample of clumps, MALT90 can also identify rare, extreme objects that 
demand ALMA's unique capabilities. While previously identified as unusual, the clump G0.25+0.02
which has the potential to form a massive Arches-like cluster but shows little star-formation, has recently gained attention as it 
may be unique in the Galaxy (Longmore et al. 2012). MALT90 and APEX maps of the kinematics and morphology 
of its dense gas reveal that it is externally heated, centrally condensed, and that it has already begun to 
fragment (Rathborne et al. submitted).  Indeed, recent ALMA cycle 0 observations reveal several tens of small, 
dense fragments ($\sim$0.1\,pc, 10$^{6}$\,\cmc) consistent with this idea (Rathborne et al. in prep).

The clumps identified by ATLASGAL and observed by MALT90 are key laboratories to study clump fragmentation and to test the 
theories of monolithic collapse and competitive accretion at high angular resolution.  Because of the large sample sizes, MALT90 
should uncover several examples of rare ``transition objects'' that bridge the gap between the major stages in clump evolution.  
Most likely, MALT90 will discover additional clumps with surprising and unanticipated properties that will need ALMA to explain.  
For reasonable assumptions about compactness, ALMA will be able to image any core detected in MALT90 at 1$''$ with 
excellent signal-to-noise.  

\section{SUMMARY}

MALT90 is a large survey that has obtained small maps in 16 molecular lines near 90 GHz toward more than $\sim$2,000 dense clumps 
across the Galaxy. The superior capabilities of the Mopra Telescope, coupled with the identification of huge numbers of dense ,
clumps from recent Galactic plane surveys, now allow, for the first time, a large-scale molecular line survey of thousands 
of dense molecular clumps.  

The primary goal of MALT90 is to characterise the physical and chemical evolution of dense, high-mass star-forming 
clumps.  The clumps were selected from the ATLASGAL point source catalog based on their IR emission so that roughly equal 
numbers of prestellar clumps, protostellar clumps, \hii\, regions and PDRs were observed. Because the maps have
excellent spatial (38\arcsec) and spectral (0.11\,\kms) resolution, the data reveal a wealth of information about the clump's 
morphologies, chemistry, and kinematics.

We have described the motivation, telescope and observing modes, 
molecular lines selected, identification of suitable high-mass star-forming clumps, the data reduction procedure and data release for the MALT90 survey.
Moreover, we have highlighted a few preliminary science results from the survey although many more are expected in the coming years.
All raw data and processed cubes and moment maps for the sources observed as part of MALT90 are publicly
available through the Australia Telescope Online Archive (ATOA; http://atoa.atnf.csiro.au/). 

MALT90 is the largest database of molecular line emission toward high-mass star-forming clumps and, thus, 
will provide a valuable dataset for studies of these regions. Moreover, the survey also provides a definitive source 
list for future ALMA observations of high-mass star-forming dense clumps.

\section*{Acknowledgments} 
JMJ gratefully acknowledges support from the US NSF grant AST-1211844. NL's postdoctoral fellowship is supported by a CONICYT/FONDECYT postdoctorado, under project no. 3130540. NL acknowledges partial support from the ALMA-CONICYT Fund for the Development of Chilean Astronomy Project 31090013. NL, LB, GG, and DM 
gratefully acknowledge support from the Center of Excellence in Astrophysics and Associated Technologies (PFB-06) and Centro de Astrof\'{i}sica FONDAP 15010003.

\end{document}